\documentclass[11pt,twoside]{article}
\usepackage{asp2004}
\usepackage{epsf}
\usepackage{graphics}
\usepackage{lscape}
\newcommand\pac{Paczy{\'n}ski }
\newcommand\ltsima{$\; \buildrel <\over\sim \;$}
\newcommand\simlt{\lower.5ex\hbox{\ltsima}}
\newcommand\gtsima{$\; \buildrel >\over\sim \;$}
\newcommand\simgt{\lower.5ex\hbox{\gtsima}}
\def\edcomment#1{\iffalse\marginpar{\raggedright\sl#1\/}\else\relax\fi}
\reversemarginpar

\begin{document}
\title{The Detection of Terrestrial Planets via Gravitational Microlensing:
Space vs. Ground-based Surveys}
\author{D. P. Bennett}
\affil{University of Notre Dame, Notre Dame, IN 46556, USA}

\begin{abstract}
I compare an aggressive ground-based gravitational microlensing survey
for terrestrial planets to a space-based survey. The Ground-based survey
assumes a global network of very wide field-of-view $\sim 2$m telescopes
that monitor fields in the central Galactic bulge. I find that such
a space-based
survey is $\sim 100$ times more effective at detecting terrestrial planets
in Earth-like orbits. The poor sensitivity of the ground-based surveys to
low-mass planets is primarily
due to the fact that the main sequence source stars are unresolved in
ground-based images of the Galactic bulge. This gives rise to
systematic photometry errors that preclude the detection of most of the
planetary light curve deviations for low mass planets.
\end{abstract}
\thispagestyle{plain}

\section{Introduction}
Gravitational microlensing can be used to detect extra-solar planets
orbiting distant stars \citep{mao-pac,gould-loeb} with relatively
large photometric signals as long as the angular planetary Einstein Ring 
radius is not much smaller than the angular size of the source star 
\citep{bennett-rhie}. For giant source stars in the Galactic bulge,
however, the signals of Earth-mass planets are largely washed out by
their large angular size, but Galactic bulge main sequence stars are
small enough to allow the detection of planets with masses as low
as $0.1 M_\oplus$. 

\begin{figure}[!ht]
\plotfiddle{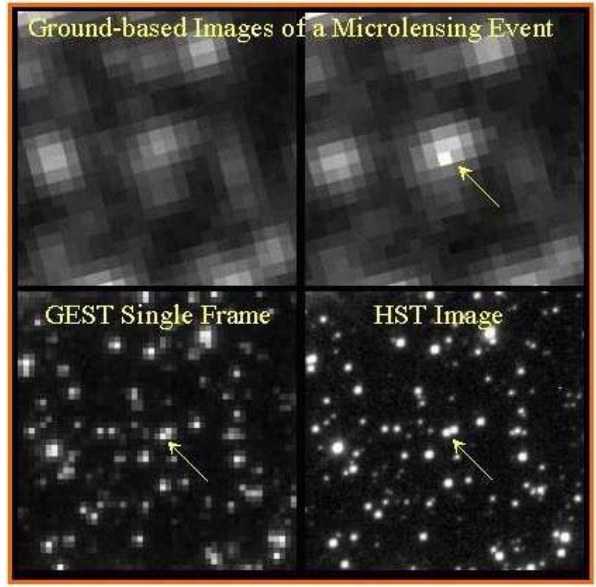}{2.60in}{0}{53}{53}{-160}{-108}
\caption{The difference between ground and space-based data for microlensing
of a bulge main sequence star is illustrated with images of microlensing
event MACHO-96-BLG-5. The two top panels are 50 min. R-band exposures
with the CTIO 0.9m telescope taken in 1" seeing at different microlensing
magnifications, and the two images on the bottom have been constructed
from HST frames. Ground-based photometry suggests that the lensed 
source star is bulge ``turn-off'' star 1-2 magnitudes brighter than
the top of the bulge main sequence, but the HST image reveals that 
the actual source star is bulge G-dwarf on the main sequence.
\label{fig-image}}

\end{figure}

This fact has led to suggestions that ground-based microlensing surveys
might be able to measure the abundance of Earth-like planets
\citep{tytler-exnps,sackett97}. However, the initial estimates of
the sensitivity of such a survey neglected the blending of main sequence
source stars in the bulge that is illustrated in Fig.~\ref{fig-image}.
The brightness of the source stars was over-estimated, and it was not
realized that the actual source stars were blended with several other
stars of similar brightness in ground-based images. More realistic
estimates of the sensitivity of the proposed ground-based extra-solar
planet searches indicated very poor sensitivity to terrestrial planets
\citep{peale-gnd_v_space}, and suggested that only a space-based survey
\citep{gest-sim} could discover a significant number of Earth-like planets.

In this paper, we simulate the most capable ground-based microlensing 
survey that could reasonably be attempted: a network of three 2-m class
wide field-of-view telescopes spanning the Globe in the Southern Hemisphere
that are dedicated to the microlensing planet search survey for four years.

\section{Simulation Details and Results}

Many of the details of our simulations have been described in
\citet{gest-sim}, but there are additional details that must be added
to describe the observing conditions at ground-based observatories. 
I assume that the microlensing survey telescopes are located at the
best existing observatory sites in Chile (Paranal), South Africa (Sutherland),
and Australia (Siding Springs). For Paranal, actual seeing and cloud cover
records are available from the ESO web site. The Sutherland site
in South Africa has conditions similar to La Silla
in Chile (Peter Martinez, private communication), so the ESO records 
for La Silla (offset by 1 year) were used as a proxy for Sutherland.
Standard airmass and wavelength correction formulae \citep{walker}
were used to convert the ESO seeing data to seeing estimates for bulge
observations in the I-band, and the sky brightness was calculated following
\citet{moon-sky}.

For the observations from Australia, the 1997-1999 seeing records from
the MACHO Project have been used since the Mt.~Stromlo Observatory and
Siding Springs have very similar seeing. To compensate for the optical
effects of the MACHO telescope, 0.8'' was subtracted in quadrature from
all the MACHO telescope seeing values. The number of clear nights was
adjusted slightly to meet the long term averages for each site as listed
in \citet{peale-gnd_v_space}.

An important input to this simulation is the systematic crowded field 
photometry errors that are assigned. The OGLE-III EWS data 
\citep{udalski} provides a set of photometry for stars with brightness
varying in a predictable way (due to microlensing) using the most accurate
crowded field photometry method devised to date \citep{isis}. By fitting
to point-source, single-lens light curves to the 2002 OGLE-III EWS 
data for stars brighter than $I < 17$, we find an assumed systematic
error of 0.7\% of the total flux at the PSF peak of each star seems to
give fit $\chi^2 \simeq 1$. The systematic errors appear to be larger
for fainter stars or for other data sets, but I adopt this as a systematic
error estimation formula.

This systematic error implies a limit to the useful size of a microlensing
planet search survey telescope. Once the systematic error dominates, it is
no longer useful to go to a larger telescope. Our simulation results yield
an optimal telescope aperture of about 2-m with a 6 sq.~deg.~field-of-view
(FOV). Larger telescopes could detect some more planets by cycling between
two or more fields, but the gain is not large because of the rapid decrease
in the microlensing rate away from the central Galactic bulge.

\begin{figure}[!ht]
\plottwo{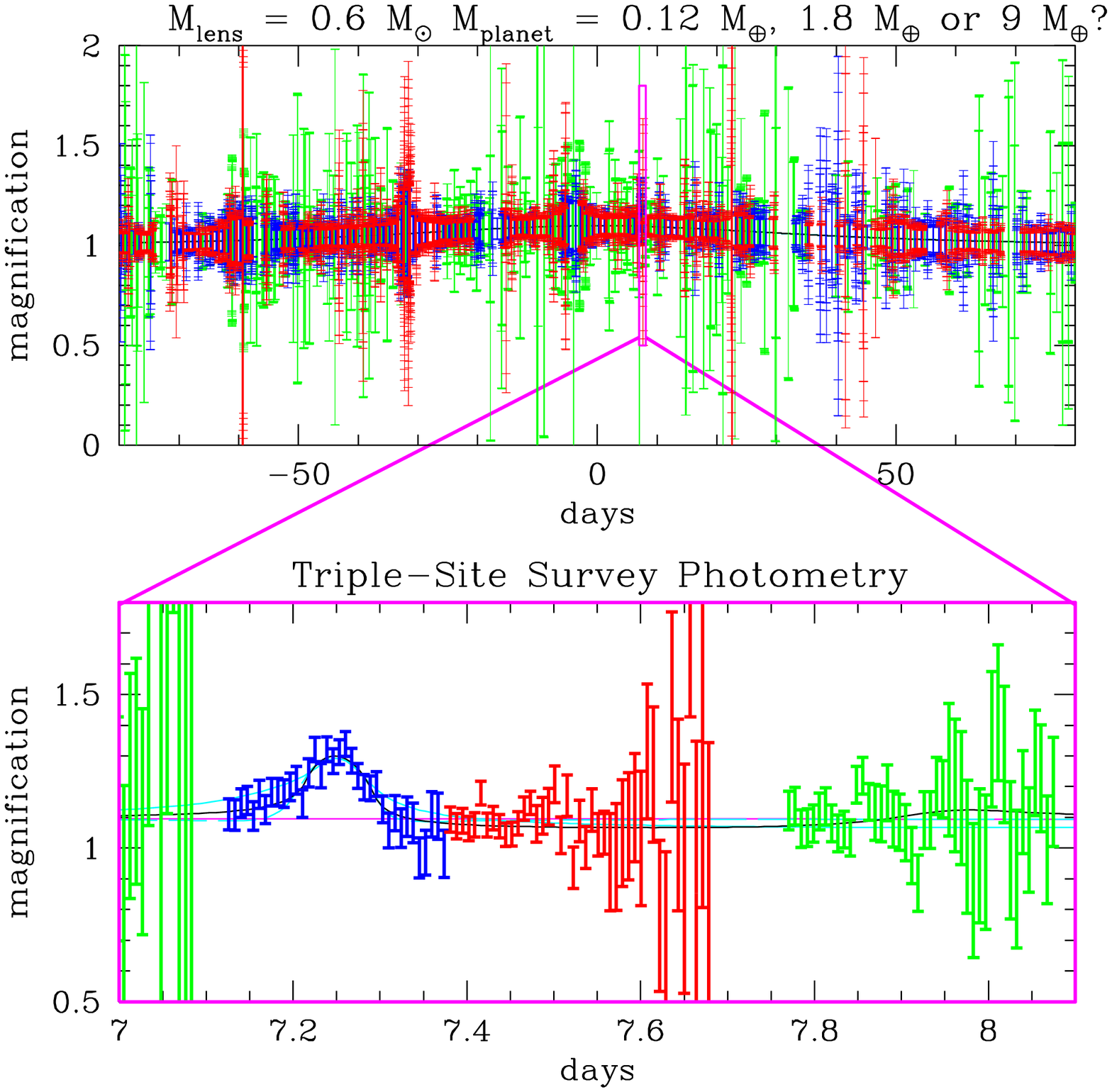}{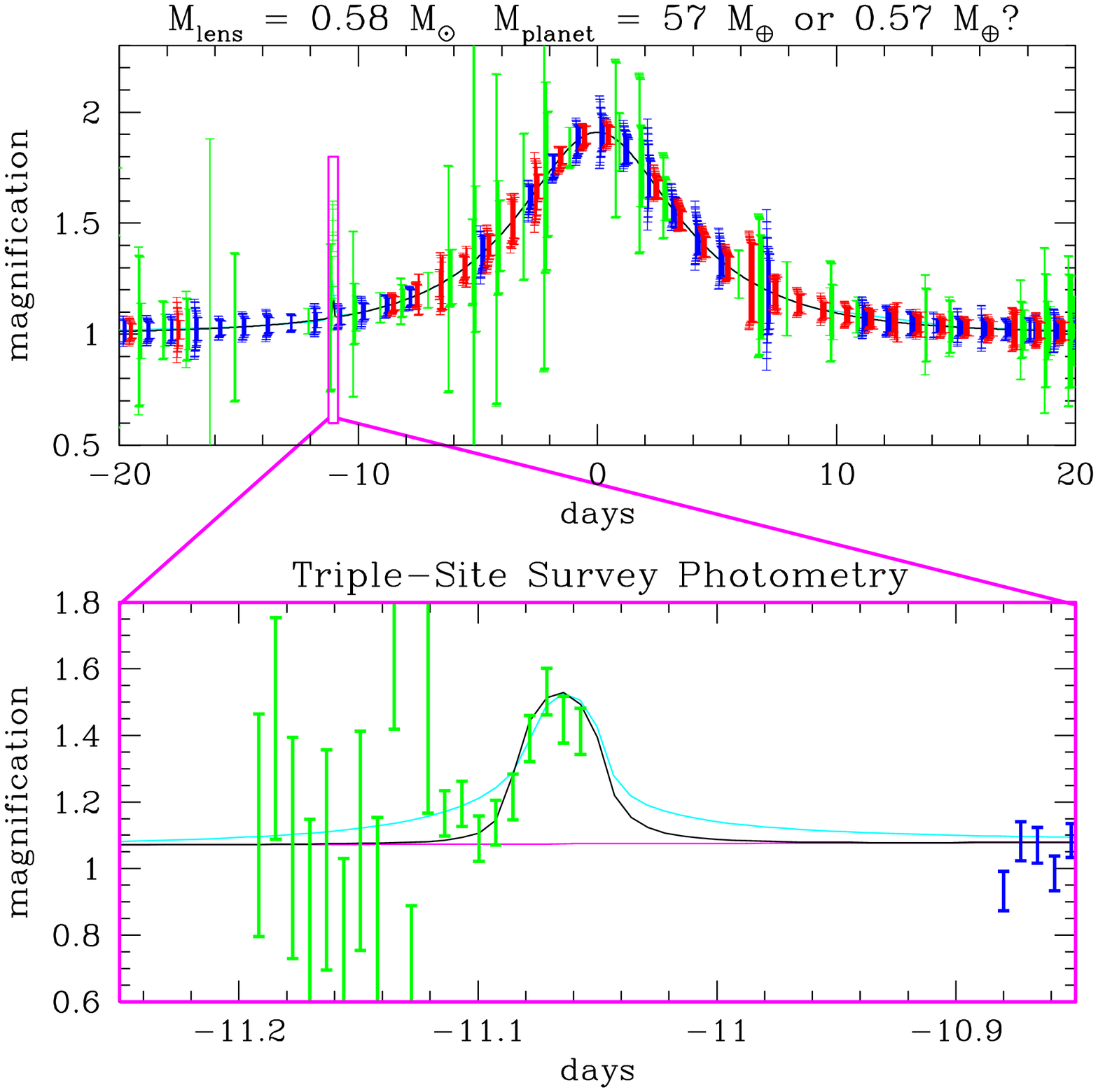}
\caption{ 
Two example light curves from our ground based
microlensing survey simulation, which allow the formal detection of the
signal of a terrestrial. Data from Chile, South Africa, and Australia are
shown represented by red, blue, and green error bars, respectively. 
The black and cyan light curves represent different microlensing models,
and the magenta curve is the best fit non-planetary model.
In both of these cases, the uncertainty in the planetary mass is more
than two orders of magnitude. The source stars for these events have
$I = 21.1$ and $I = 20.8$, respectively.
\label{fig-bad-lc1}}
\end{figure}

\begin{figure}[!ht]
\plottwo{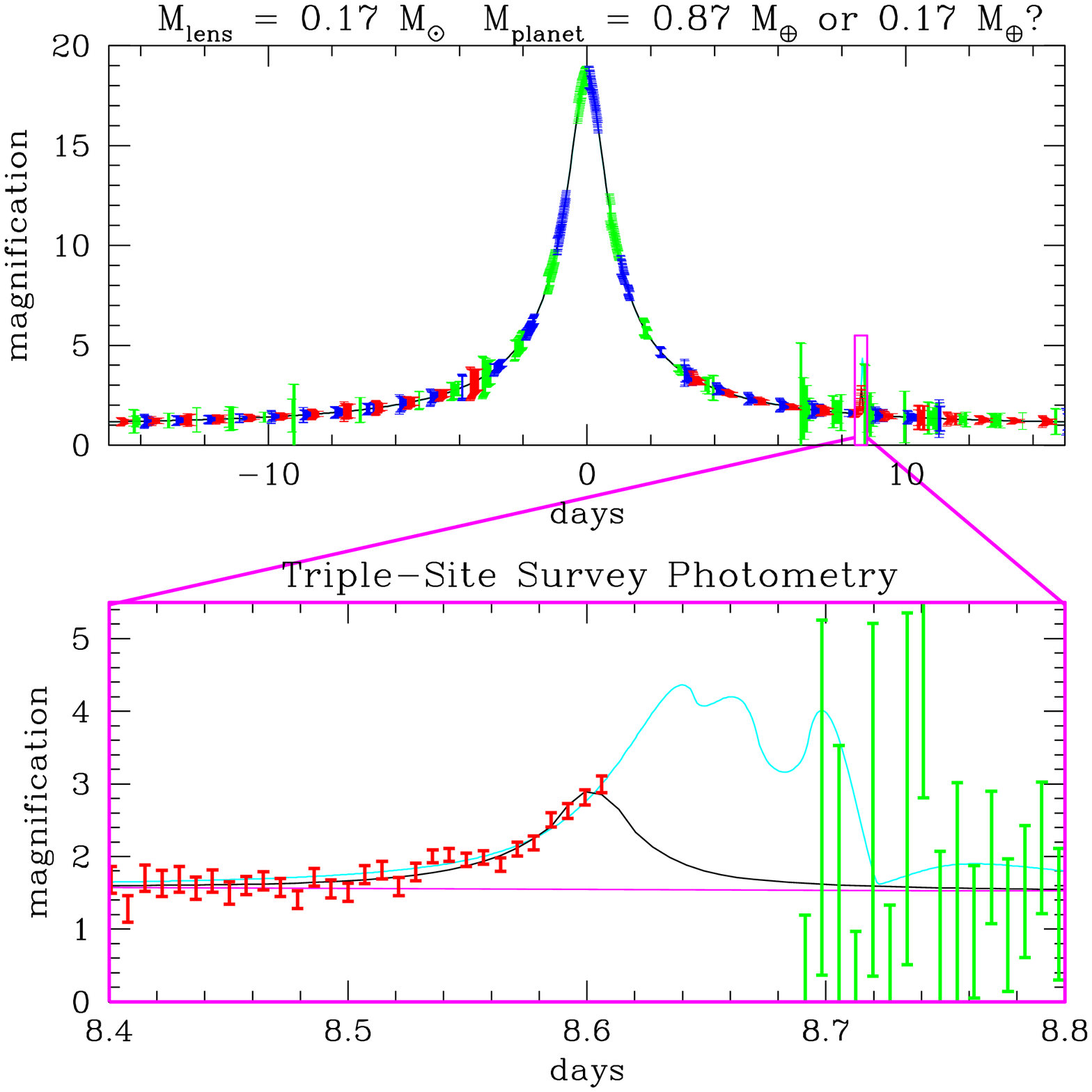}{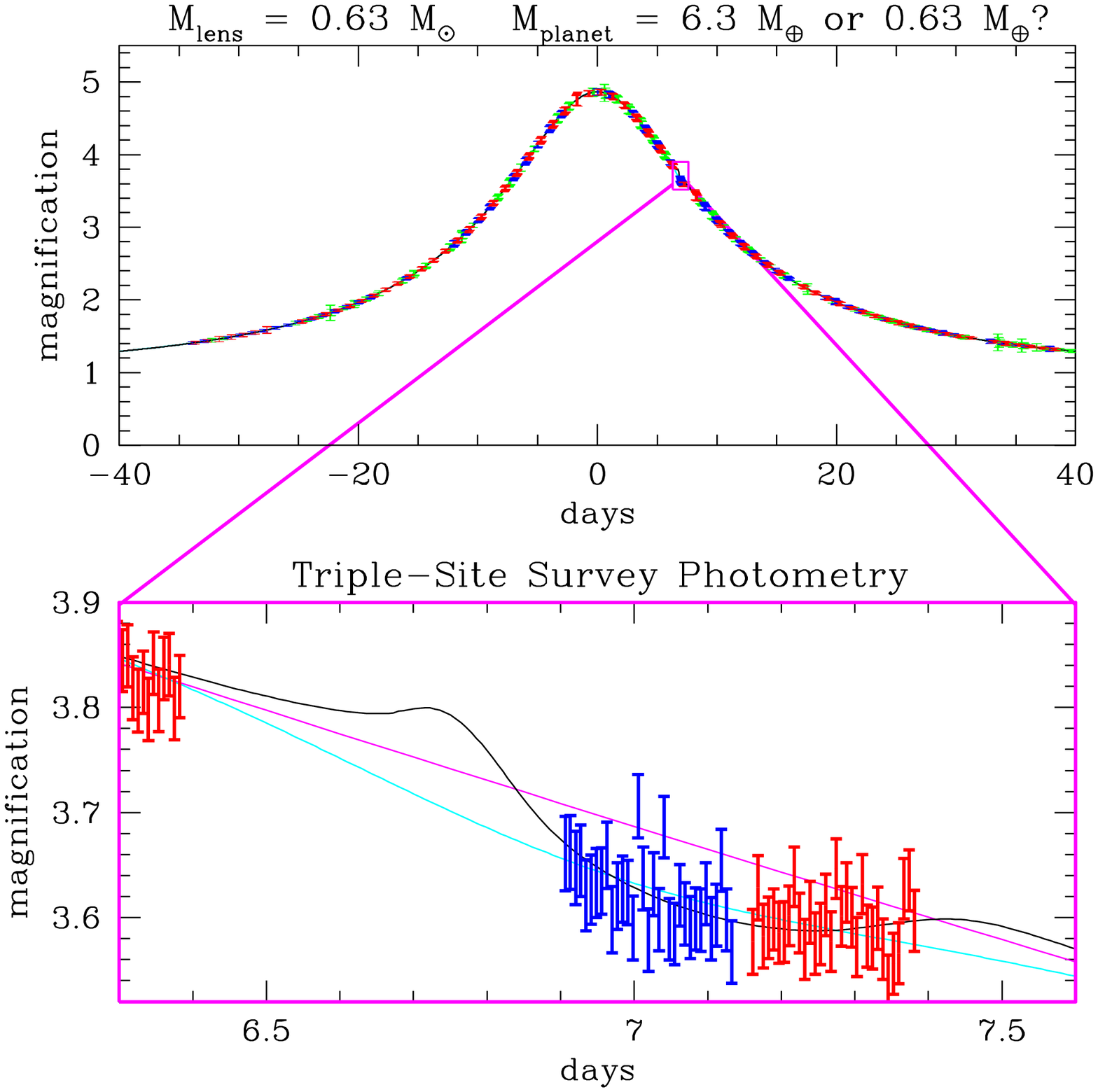}
\caption{ 
Two additional simulated light curves for events with detected light
curve deviations due to terrestrial planets where the light curve coverage
or quality is not sufficient to determine the planetary mass. The source
stars for these events have $I = 22.0$ and $I = 18.2$, respectively.
\label{fig-bad-lc2}}
\end{figure}

Fig.~\ref{fig-bad-lc1} shows an example of two light curves that would
be considered planet detections by the criteria of \citet{gest-sim} for
a space-based survey (a $\Delta\chi^2 > 160$ improvement for the planetary
lensing model over the best single lens model). But the uneven data 
quality due to seeing variations and cloud cover imply that the 
planetary parameters cannot be constrained. Thus, while the planetary 
signal is detectable for these events, terrestrial planets cannot
be discovered with data like this. The situation is only slightly better
for the events shown in Fig.~\ref{fig-bad-lc2}, where the poor seeing 
in Australia, and
low S/N of the planetary deviation conspire to prevent determinations
of the planetary masses.

\begin{figure}[!ht]
\plottwo{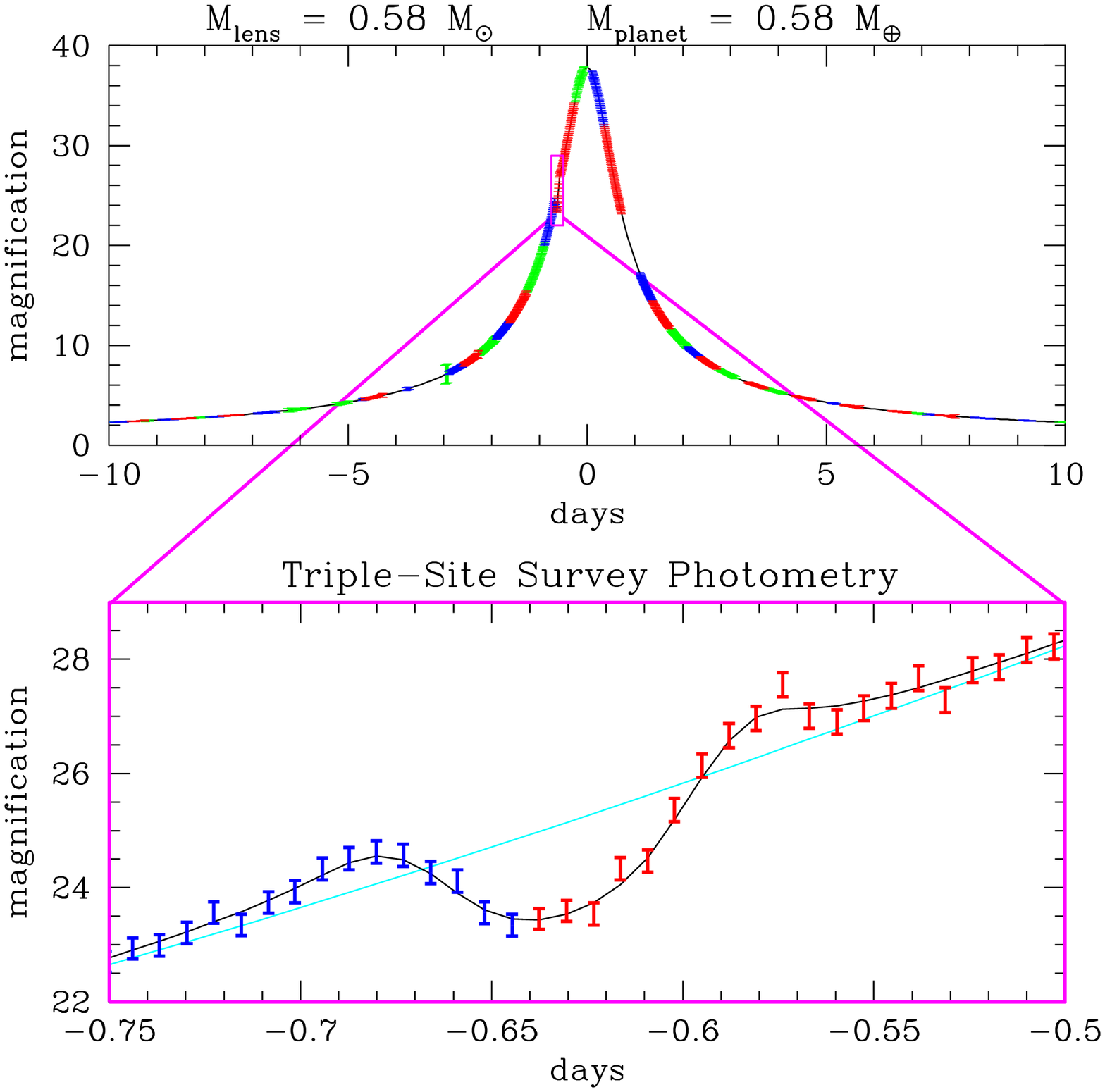}{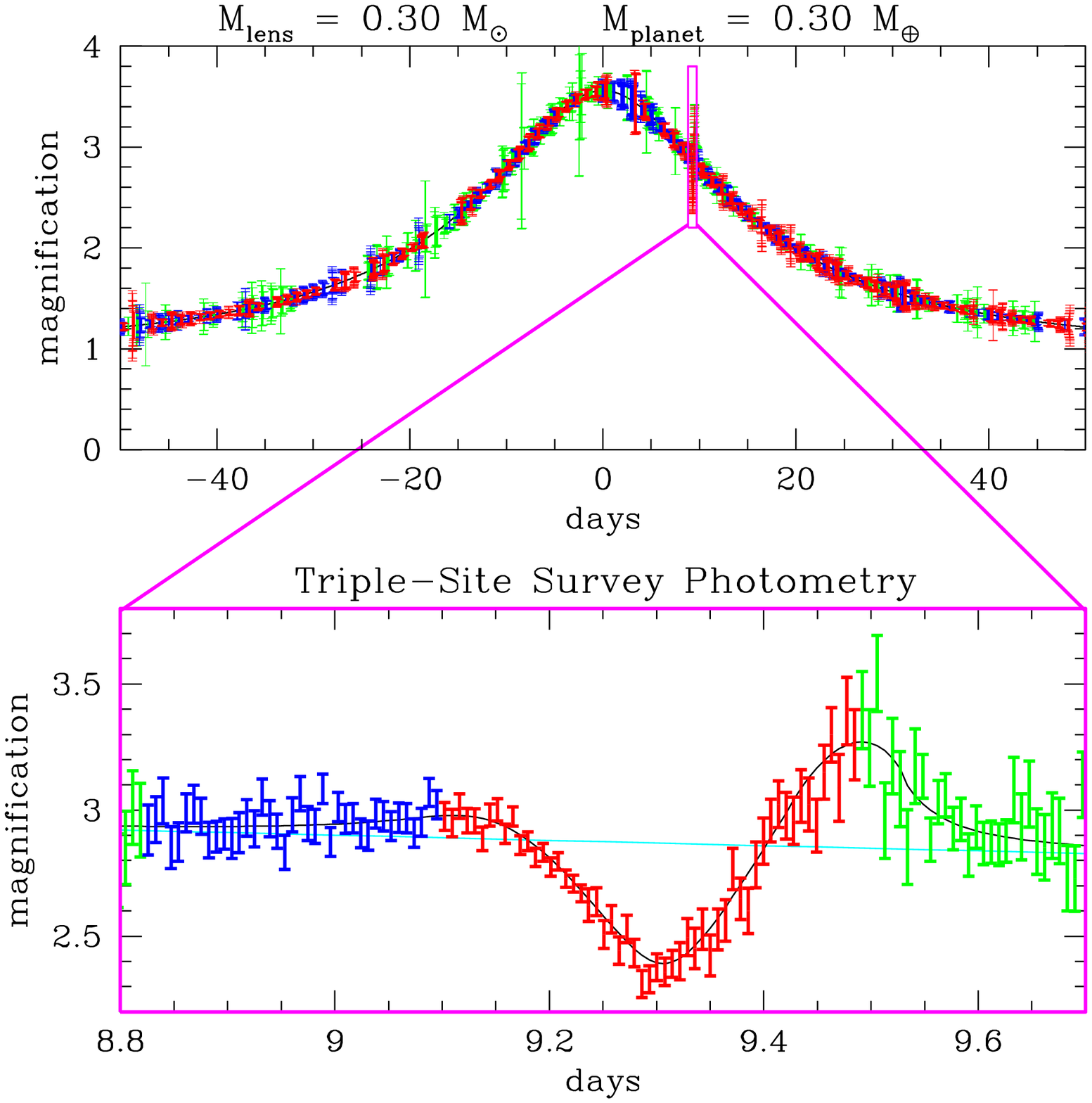}
\caption{ 
Examples of light curves from our simulations where the light curve 
coverage and photometric accuracy allows for accurate determination
of the terrestrial planetary parameters. These are both examples of 
simulated planetary discoveries. The source stars for these events have
$I = 20.5$ and $I = 20.0$, respectively.
\label{fig-good-lc}}
\end{figure}

Of course, sometimes a planetary signal will be seen during good 
observing conditions, yielding light curves that can accurately
determine the planetary parameters. This is the case for the two
events shown in Fig.~\ref{fig-good-lc}. Generally, such events have
planetary deviations that occur at high magnification, such as the
event on the left or have large amplitude deviations like the event
on the right.

\begin{figure}[!ht]
\plottwo{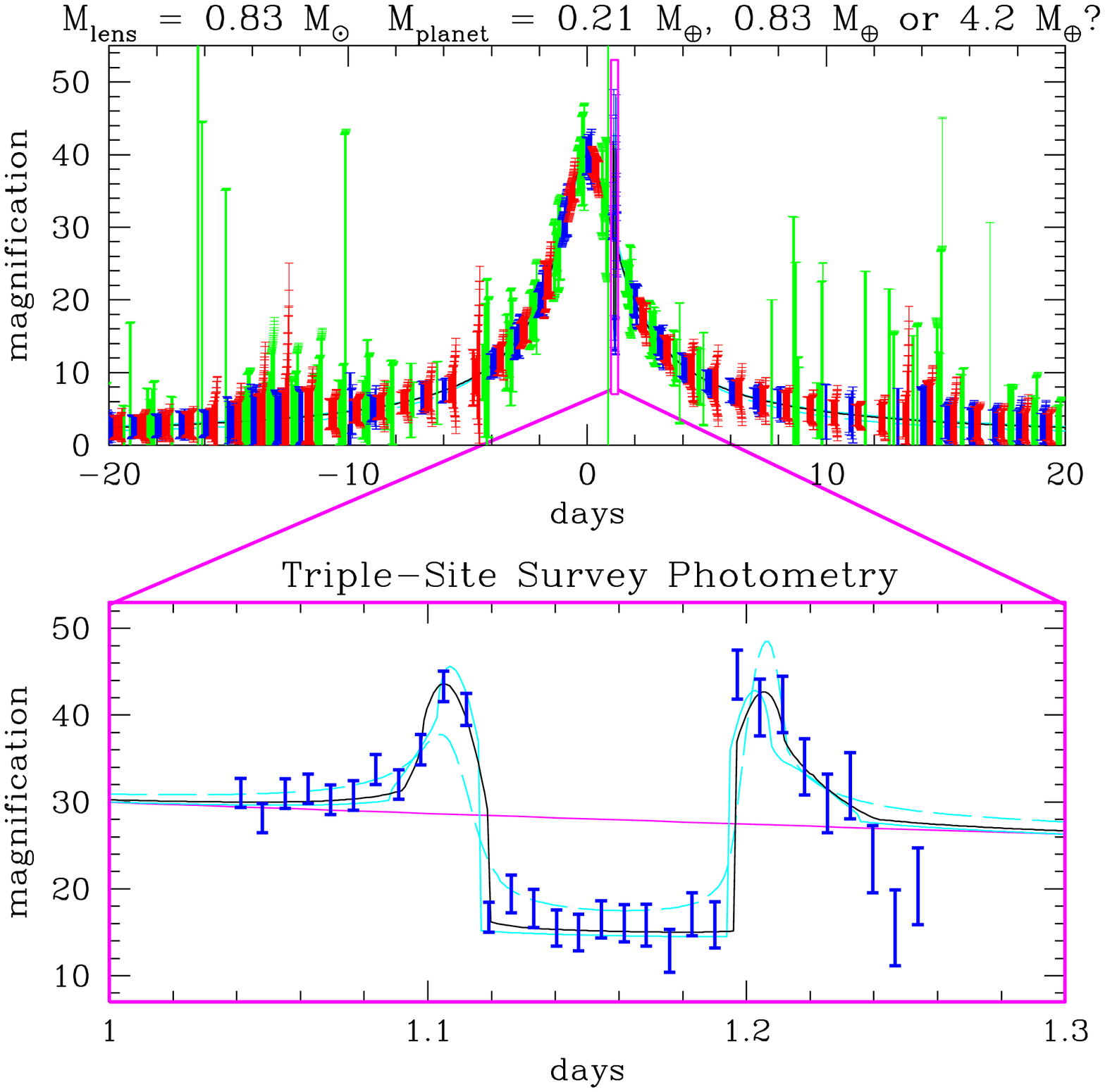}{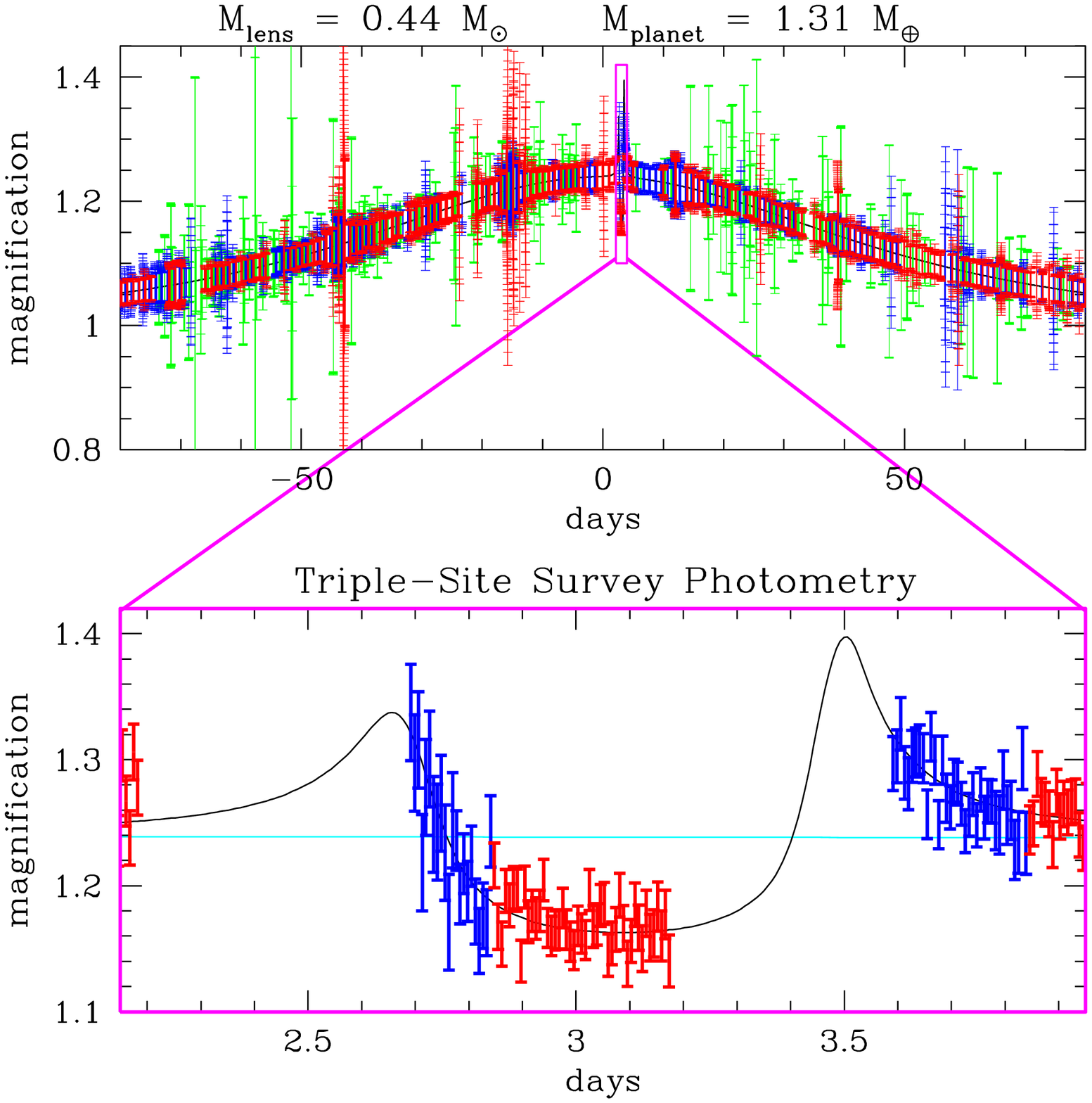}
\caption{ 
Examples of simulated light curves with detected terrestrial planet signals,
where the planetary parameters can formally be determined. The source stars
for these events have $I = 24.6$ and $I = 19.3$, respectively.
\label{fig-fair-lc}}
\end{figure}

Intermediate between the poorly covered planetary detections shown
in Figs.~\ref{fig-bad-lc1}--\ref{fig-bad-lc2} and the well covered
planetary discoveries shown in Figs.~\ref{fig-good-lc} are the light
curves shown in Fig.~\ref{fig-fair-lc}. In both cases, the planetary
parameters can be formally determined. However for the event on the
left, the source star brightness is only $I = 24.6$. If the
photometry errors are not random, as is likely for systematic errors,
then the brightness of the source star cannot be determined from the
light curve fit. Hence, the planetary parameters will be poorly constrained.
For the light curve on the right, the parameters can be determined
despite the relatively poor coverage of the deviation because the
deviation shape contains redundant information. However, the poor
light curve coverage means that the redundancy cannot be used to
confirm the interpretation of the event. This is a serious drawback
because the planet discovery cannot be confirmed in any other way.

It is clear from Figs.~\ref{fig-bad-lc1}-\ref{fig-good-lc} it is sensible
to consider two different categories of detected planetary events:
``detected'' planets and ``discovered'' planets. The ``detected'' category
includes all planets with planetary signals that provide at least a
$\Delta\chi^2 \geq 160$ improvement over the best single lens fit.
Clearly, the events shown in Fig.~\ref{fig-good-lc} should be in the 
``discovered'' while those shown in Figs.~\ref{fig-bad-lc1} and 
\ref{fig-bad-lc2} should not be in this category. We define the planet
``discovered'' category with the following criteria. Each planetary 
light curve deviation is split up into 1-3 separate deviation regions
of positive or negative magnification with respect to the single lens
curve with the same parameters. The first and last regions are considered
be begin and end when the deviation reaches the with 10\% of the maximum
planetary deviation or 0.3\% of single lens magnification. For a planet 
``discovery'', each the observations must cover at least 40\% of each
planetary deviation region, and at least 60\% of each total deviation.
All measurements are considered to contribute the light curve coverage
unless their error estimates are larger than one third of the maximum 
planetary deviation. Also, because a significant component of the error
estimates is a systematic error, we also require that each ``discovery"
light curve have at least one measurement that detected the stellar
microlensing event by $\geq 10\sigma$ and one measurement that detects
the planetary deviation by $\geq 5\sigma$. With these definitions, about
one third of the detected planets pass the planet discovery threshold
including both the events in Fig.~\ref{fig-good-lc} and the light curve
on the left side of Fig.~\ref{fig-fair-lc}.

\begin{figure}[!ht]
\plotfiddle{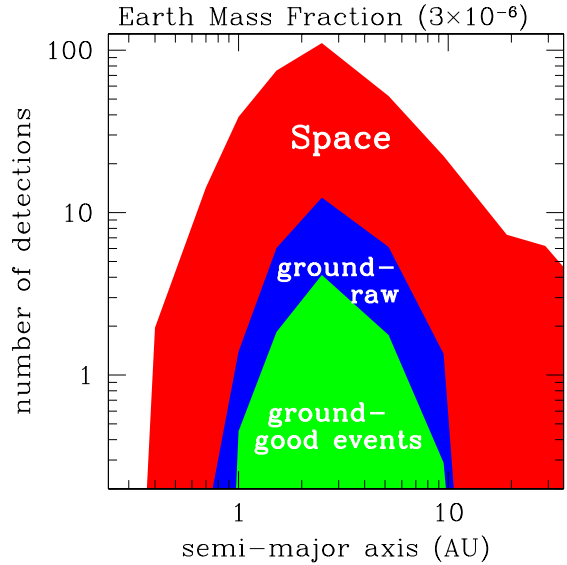}{2.6in}{0}{40}{40}{-126}{-72}
\caption{
The number of expected planetary detections as a function of orbital semi-major
axis is compared for a space-based survey and a 3-site ground based survey.
Planets with the Earth's mass fraction of $\epsilon = 3\times 10^{-6}$ are
considered.
The green region indicates the events detected from the ground that are
well characterized, while the blue region indicates events which have a
detectable signal which are not well characterized, as described in the
text. The red region is the predicted sensitivity of the proposed GEST
space mission according to the simulations of \citet{gest-sim}.
\label{fig-det_sep}}
\end{figure}

Fig.~\ref{fig-det_sep} compares the sensitivity of a 3-site network of 
2m-class ground-based microlensing planet search telescopes vs.~a 
space-based microlensing planet search telescope that would be appropriate
for NASA's Discovery Program \citep{gest-spie}. At present, the cost cap
for a Discovery mission is \$360M, and the ground-based network that
I have simulated would be roughly one fifth that cost, including all 
software development and operational expenses. The ground-based survey
does the best at separations of 2--3 AU where the planet is near the
Einstein Ring and is detectable in high magnification events. The ground-based
survey would have about 30 times fewer planet discoveries and 10 times fewer
raw detections of terrestrial planets at these separations. At 1 AU, the
advantage of space for planet discoveries is a factor of 100.

\begin{figure}[!ht]
\plottwo{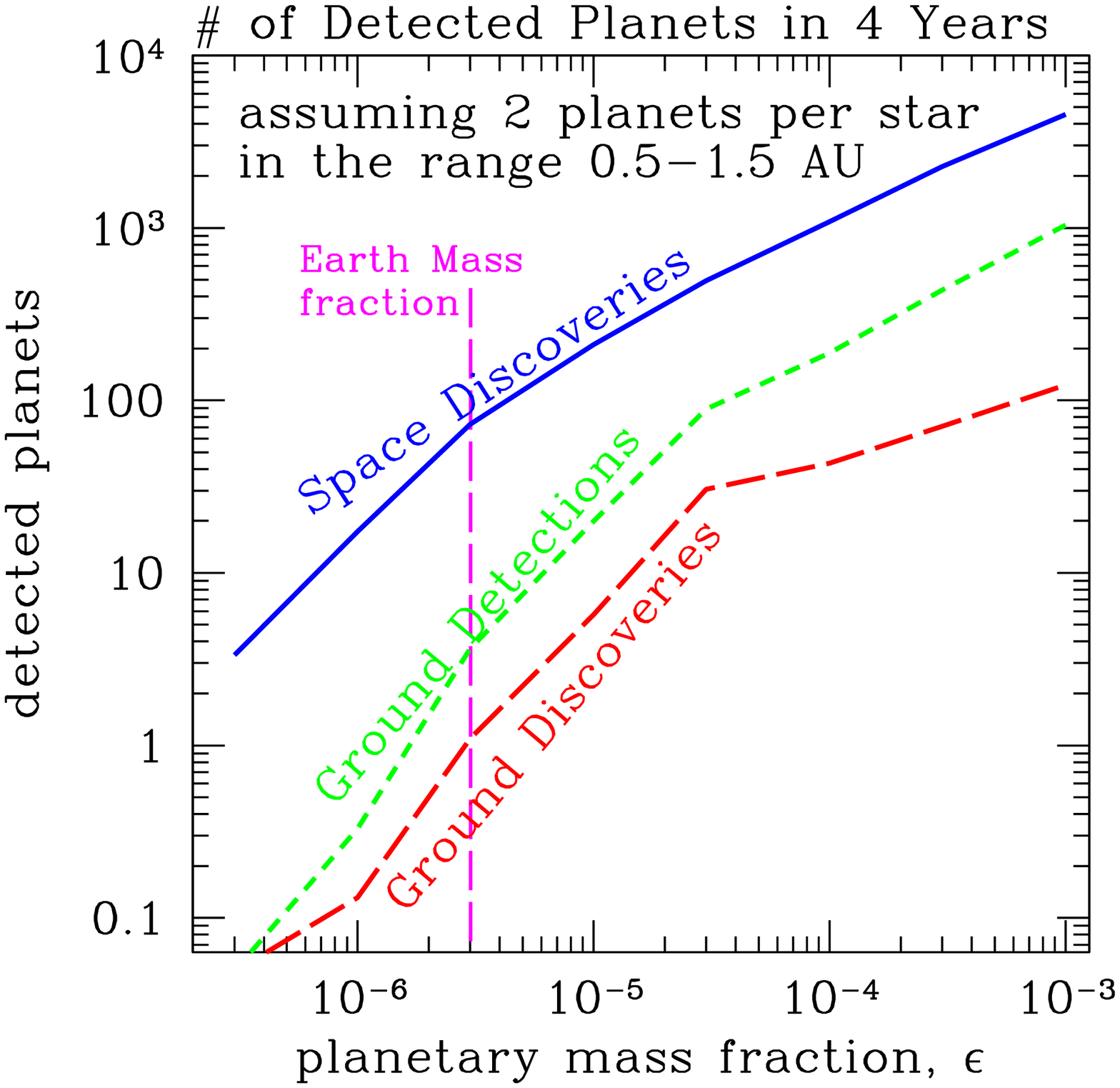}{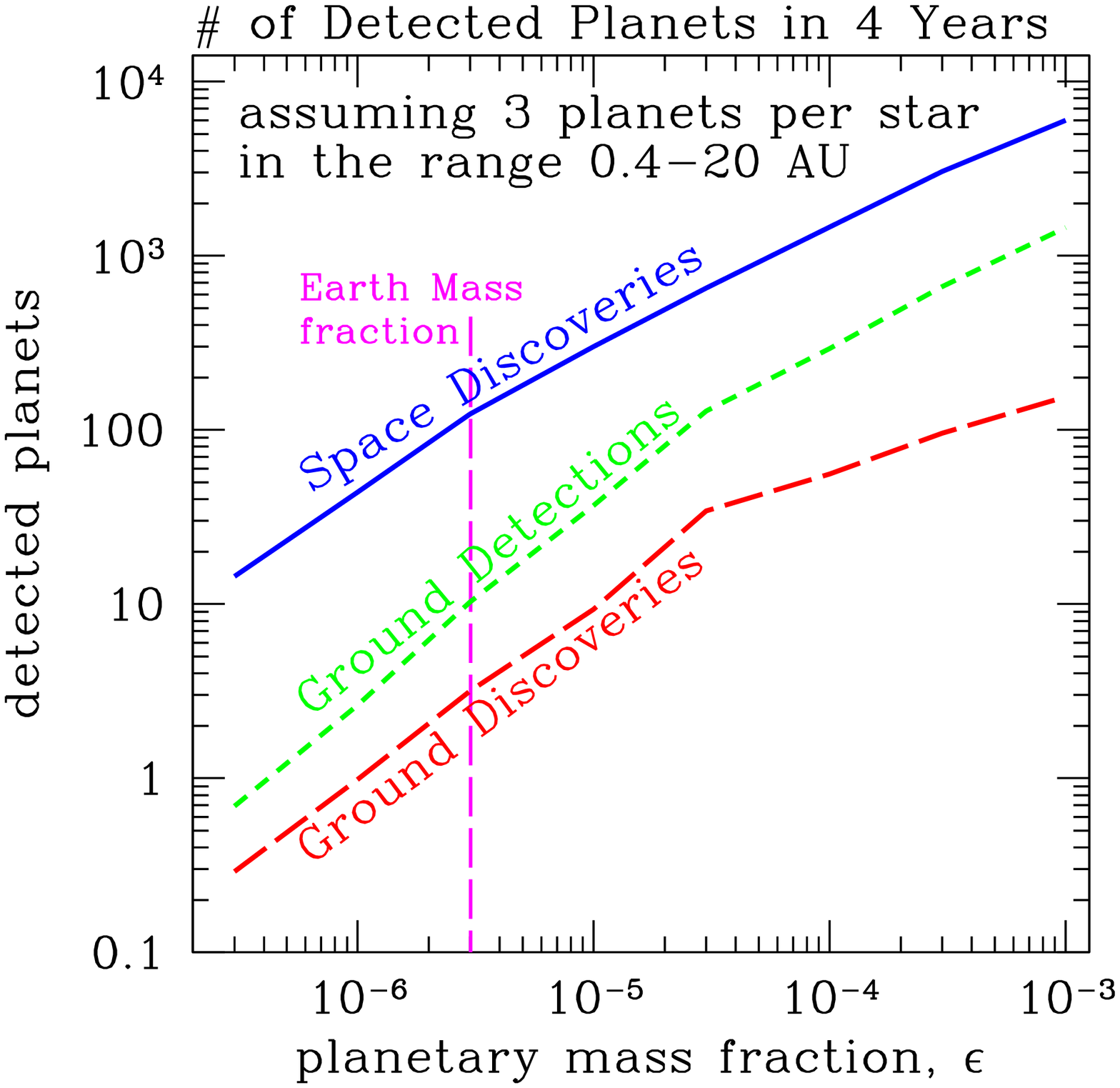}
\caption{
The number of planet discoveries and detections as a function of the
planetary mass fraction, $\epsilon$, is compared for a 3-site ground-based
wide FOV telescope survey and a dedicated space mission. The left hand
panel shows the comparison for planets in Earth-like orbits, while the
right hand panel shows all planets.
\label{fig-det_eps}}
\end{figure}

The number of planet detections as a function of planetary mass fraction
are shown in Fig.~\ref{fig-det_eps} for planets in Earth-like orbits and
planets in all orbits. The left hand panel indicates that even if Earth-like
planets in Earth-like orbits are quite abundant, the 3-site ground-based
survey would only expect to find one with good light curve coverage.

The ground-based results in Fig.~\ref{fig-det_eps} are somewhat misleading
for $\epsilon \geq 3\times 10^{-5}$ because these events can generally be
detected with giant source stars, which have not been included in this
simulation. These would add significantly to the ground-based detections,
but would add little to the space-based detections. So, for planets with
mass fractions about ten times that of the Earth, the ground-based
survey may have an advantage.

\begin{figure}[!ht]
\plotfiddle{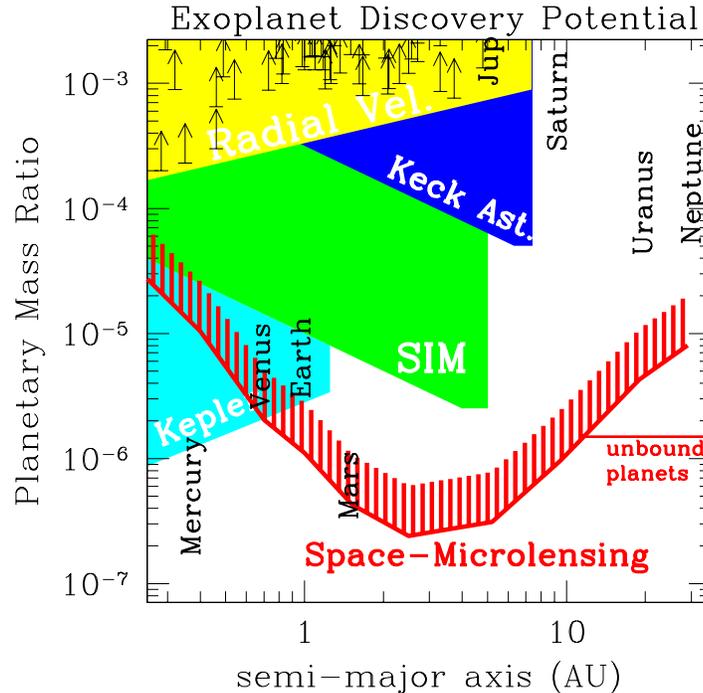}{3.15in}{0}{50}{50}{-162}{-100}
\caption{
The sensitivity of a space-based microlensing survey
is plotted as a function of planetary mass
fraction, $\epsilon$, and orbital semi-major axis.
The yellow region gives the sensitivity of a 20-year radial
velocity program on the Keck Telescope assuming a detection threshold of
10 m/sec, and the blue region indicates the
sensitivity of a 10-year interferometric astrometry program with a
30$\,\mu$as detection threshold.
The green and cyan regions indicate the sensitivities of the SIM and Kepler
missions which are in development. The location of our Solar
System's planets and some of the extra-solar planets detected
by radial velocities are shown. 
The horizontal lines indicate the sensitivity to free-floating
planets since the more distant planets can sometimes be detected without
seeing a microlensing signal from their star.
Space-based microlensing clearly
complements Kepler with its sensitivity to low-mass terrestrial
planets at separations $\geq 0.7\,$AU.
\label{fig-n_vs_sep}}
\end{figure}

\section{Conclusions}

I have carried out detailed simulations of the most capable ground-based 
microlensing terrestrial planet search program that could plausibly 
be attempted. The simulated survey employs three 2m class telescopes 
spanning the Globe in the Southern Hemisphere which observe the Galactic 
bulge whenever possible for a period of 4 years. Such a network is 
about 100 times less sensitive than a space-based survey which might 
only cost 5 times as much. So, without some dramatic improvement in
ground-based crowded field photometry, it will not be possible to 
conduct a microlensing survey to determine the abundance of terrestrial
planets in the inner Galaxy. 

For a gravitational census of terrestrial planets, a space-based survey 
will be required \citep{gest-sim,gest-spie}. As Fig.~\ref{fig-n_vs_sep} 
indicates, such a mission would complete the survey for terrestrial 
planets that will be started by the Kepler mission \citep{kepler} which
is sensitive to Earth-like planets at separations $\la 1\,$AU using the
transit method. A space-based microlensing survey would overlap Kepler
with sensitivity at $\sim 1\,$AU, and extend sensitivity to terrestrial
planets to all separations including free-floating planets, which may
have been ejected from their parent stars during the planetary formation
process.

\end{document}